\documentstyle[prl,aps]{revtex}
\twocolumn

\begin{document}
\author{Jian Qi Shen $^{1,}$$^{2}$}
\address{$^{1}$  Centre for Optical
and Electromagnetic Research, State Key Laboratory of Modern
Optical Instrumentation, \\Zhejiang University,
Hangzhou Yuquan 310027, P.R. China\\
$^{2}$ Zhejiang Institute of Modern Physics and Department of
Physics, Zhejiang University, Hangzhou 310027, P.R. China}
\date{\today }
\title{Negative optical refractive index resulting from a moving regular medium}
\maketitle

\begin{abstract}
The optical refractive index of a moving regular medium is
calculated by using the Lorentz transformation in this note. It is
shown that in some velocity region of medium moving relative to
the initial frame K, the moving medium may possess a negative
index of refraction measured by the observer fixed at the frame of
reference K.

\end{abstract}

More recently, a class of artificial composite metamaterials
(termed ``left-handed media''), which possess the {\it negative}
indices of optical refraction, attract extensive attention in
various areas such as applied electromagnetism, classical optics,
materials science as well as condensed matter
physics\cite{Smith,Klimov,Ziolkowski2,Kong,Pendryprl,Zhang,Jianqi,Shen}.
Pendry {\it et al.} suggested that such left-handed materials can
be designed and fabricated by using both the {\it array of long
metallic wires} (ALMWs)\cite{Pendry2} and the {\it split ring
resonators} (SRRs)\cite{Pendry1,Maslovski}. A combination of the
two structures yields a left-handed medium. Apart from such
artificial fabrications, does there exist another alternative to
the negative index of refraction? In this note, I will demonstrate
that a moving regular medium with a velocity in a suitable region
(associated with its rest refractive index $n$)\footnote{The rest
refractive index is just the one measured by the observer fixed at
the regular/ordinary medium. In what follows, we will derive the
relativistic transformation for the optical refractive index
(tensor) of a moving medium.} possesses a negative index of
refraction, which may be a physically interesting effect and might
deserve further discussion.

Suppose we have a regular anisotropic medium with the rest
refractive index tensor $\hat{n}$, which is moving relative to an
inertial frame K with speed ${\bf v}$ in the arbitrary directions.
The rest refractive index tensor $\hat{n}$ can be written as $
\hat{n}={\rm diag}\left[n_{1}, n_{2}, n_{3}\right] $ with
$n_{i}>0, i=1,2,3$. Thus the wave vector of a propagating
electromagnetic wave with the frequency $\omega$ reads
$
{\bf k}=\frac{\omega}{c}\left(n_{1}, n_{2}, n_{3}\right)
\label{eqq1}
$
 measured by the observer fixed at this medium. In this sense,
one can define a 3-D vector ${\bf n}=\left(n_{1}, n_{2},
n_{3}\right)$, and the wave vector ${\bf k}$ may be rewritten as
${\bf k}={\bf n}\frac{\omega}{c}$. Now we analyze the phase
$\omega t-{\bf k}\cdot{\bf x}$ of the above time-harmonic
electromagnetic wave under the following Lorentz transformation
\begin{equation}
{\bf x}'=\gamma\left({\bf x}-{\bf v}t\right), \quad
t'=\gamma\left(t-\frac{{\bf v}\cdot{\bf x}}{c^{2}}\right),
\label{eq1}
\end{equation}
where $\left({\bf x}', t'\right)$ and $\left({\bf x}, t\right)$
respectively denote the spacetime coordinates of the initial frame
K and the system of moving medium, the spatial origins of which
coincide when $t=t'=0$. Here the relativistic factor
$\gamma=\left(1-\frac{v^{2}}{c^{2}}\right)^{-\frac{1}{2}}$. Thus
by using the transformation (\ref{eq1}), the phase $\omega t-{\bf
k}\cdot{\bf x}$ observed inside the moving medium may be rewritten
as the following form by using the spacetime coordinates of K
\begin{equation}
\omega t-{\bf k}\cdot{\bf x}=\gamma\omega\left(1-\frac{{\bf
n}\cdot{\bf v }}{c}\right)t'-\gamma\omega\left(\frac{{\bf n
}}{c}-\frac{{\bf v}}{c^{2}}\right)\cdot{\bf x}',
\end{equation}
the term on the right-handed side of which is just the expression
for the wave phase in the initial frame K. Hence, the frequency
$\omega'$ and the wave vector ${\bf k}'$ of the observed wave we
measure in the initial frame K are given
\begin{equation}
\omega'=\gamma\omega\left(1-\frac{{\bf n}\cdot{\bf v }}{c}\right),
\quad        {\bf k}'=\gamma\omega\left(\frac{{\bf n
}}{c}-\frac{{\bf v}}{c^{2}}\right),             \label{eq2}
\end{equation}
respectively.

To gain some insight into the meanings of the expression
(\ref{eq2}), let us consider the special case of a boost (of the
Lorentz transformation (\ref{eq1})) in the
$\hat{x}_{1}$-direction, in which the medium velocity relative to
K along the positive $\hat{x}_{1}$-direction is $v$. In the
meanwhile, we assume that the wave vector of the electromagnetic
wave is also parallel to the positive $\hat{x}_{1}$-direction. If
the {\it rest} refractive index of the medium in the
$\hat{x}_{1}$-direction is $n$, its (moving) refractive index in
the same direction measured by the observer fixed at the initial
frame K is of the form
\begin{equation}
n'=\frac{ck'}{\omega'}=\frac{n-\frac{v}{c}}{1-\frac{nv}{c}},
\label{eq3}
\end{equation}
which is a relativistic formula for the addition of ``refractive
indices'' ($-\frac{v}{c}$ provides an effective index of
refraction).

Note that here $n'$ observed in the frame K may be negative. If,
for example, when $n>1$, the medium moves at
\begin{equation}
c>v>\frac{c}{n}                           \label{eq5}
\end{equation}
(and in consequence, $k'>0$, $\omega'<0$); or when $0<n<1$, the
medium velocity with respect to K satisfies
\begin{equation}
c>v>nc                                         \label{eq6}
\end{equation}
(and consequently, $k'<0$, $\omega'>0$), then $n'$ is negative.
The former case where $k'>0$, $\omega'<0$ is of physical interest.
In the paper\cite{Shen}, it was shown that the photon in the
negative refractive index medium behaves like its {\it
anti-particle}, which, therefore, implies that we can describe the
wave propagation in both left- and right- handed media in a
unified way by using a complex vector field\cite{Shen}.

In the above, even though we have shown that in some certain
velocity regions, the moving medium possesses a negative $n'$,
such a moving medium cannot be surely viewed as a left-handed
medium. For this point, it is necessary to take into account the
problem as to whether the vector {\bf {k}}, the electric field
{\bf {E}} and the magnetic field {\bf {H}} of the electromagnetic
wave form a left-handed system or not\footnote{This requirement is
the standard definition of a left-handed medium. Apparently, it is
seen that the two concepts {\it left-handed medium} and {\it
negative refractive index medium} is not completely equivalent to
each other.}. For simplicity and without the loss of generality,
we choose the electric and magnetic fields of the electromagnetic
wave in the medium system as

\begin{equation}
{\bf E}=\left(0, E_{2}, 0\right),   \quad   {\bf
B}=\left(0,0,B_{3}\right).
\end{equation}
In the regular medium (right-handed medium with $n>0$), the wave
vector (along the positive $\hat{x}_{1}$-direction with the
modulus $k$) and such ${\bf E}$ and ${\bf B}$ (or ${\bf H}$) form
a right-handed system\footnote{Thus if $E_{2}>0$ is chosen to be
positive, then $B_{3}$ is also positive. In what follows, we will
adopt this case ({\it i.e.}, $E_{2}>0$ and $B_{3}>0$) without the
loss of generality.}. In the frame of reference K, according to
the Lorentz transformation, one can arrive at

\begin{equation}
E'_{2}=\gamma\left(E_{2}-vB_{3}\right)       \quad
B'_{3}=\gamma\left(B_{3}-\frac{v}{c^{2}}E_{2}\right). \label{eq7}
\end{equation}
For convenience, here we will think of the medium as an isotropic
one, {\it i.e.}, the refractive index $n'$, permittivity
$\epsilon'$ and the permeability $\mu'$ of the moving medium are
all the scalars rather than the tensors. First we consider the
case of Eq.(\ref{eq6}) where $k'<0$, $\omega'>0$. In order to let
the moving medium with a negative index of refraction be a real
left-handed one, the wave vector ($k'<0$), $E'_{2}$ and $H'_{3}$
of the electromagnetic wave measured in K should form a
left-handed system. Because of $k'<0$ and
$B'_{3}=\mu'\mu_{0}H'_{3}$ with $\mu_{0}$ being the magnetic
permeability in a vacuum, if $\mu'$ is negative\footnote{In the
left-handed medium, the electric permittivity and the magnetic
permeability are simultaneously negative.}, then only the case of
$E'_{2}<0$ and $B'_{3}>0$ (and vice versa) will lead to the
left-handed system\footnote{Since $k'<0$, in order to form a
left-handed system, the signs of $E'_{2}$ and $H'_{3}$ should not
be opposite to each other. So, the signs of $E'_{2}$ and $B'_{3}$
should be opposite due to $\mu'<0$.} formed by the wave vector
($k'<0$), $E'_{2}$ and $H'_{3}$. So, it follows from
Eq.(\ref{eq7}) that when the medium velocity relative to K is in
the range
\begin{equation}
\frac{c^{2}B_{3}}{E_{2}}<v<\frac{E_{2}}{B_{3}}      \quad
\left({\rm if}  \quad  E^{2}_{2}>c^{2}B^{2}_{3}\right),
\label{eq8}
\end{equation}
or
\begin{equation}
\frac{c^{2}B_{3}}{E_{2}}>v>\frac{E_{2}}{B_{3}}      \quad
\left({\rm if}  \quad  E^{2}_{2}<c^{2}B^{2}_{3}\right),
\label{eq9}
\end{equation}
then the wave vector, electric field and magnetic field measured
in K will truly form a left-handed system.

In conclusion, if the medium velocity satisfies both
Eq.(\ref{eq6}) and Eq.(\ref{eq8}) or (\ref{eq9}), then the moving
medium will be a left-handed one observed from the observer fixed
in the initial frame K.

As far as the case of Eq.(\ref{eq5}) where $k'>0$, $\omega'<0$ is
concerned, it is also possible for the moving material to become a
left-handed medium. But the case of $\omega'<0$ has no the
practical counterpart (at least in the real situation where the
artificial composite materials is designed). So, we would not
further discuss the case of $\omega'<0$.

Although the subject of the present note seems to be somewhat
trivial, it is helpful for understanding the properties of wave
propagation (such as causality problem, the problem of energy
propagating outwards and backwards from source\cite{Smith2}, {\it
etc.}) in the left-handed medium.

\textbf{Acknowledgements}  This work was supported partially by
the National Natural Science Foundation of China under Project No.
$90101024$.


\begin{references}
\bibitem{Smith} D.R. Smith, W.J. Padilla, D.C. Vier{\it et
al.}, Phys. Rev. Lett. 84 (2000) 4184;  R.A. Shelby, D.R. Smith,
and S. Schultz, Science 292 (2001) 77.

\bibitem{Klimov} V.V. Klimov, Opt. Comm. 211 (2002) 183.


\bibitem{Ziolkowski2} R.W. Ziolkowski, Phys. Rev. E 64 (2001)
056625.

\bibitem{Kong} J.A. Kong, B.L. Wu, and Y. Zhang, Appl. Phys. Lett.
80 (2002) 2084; N. Garcia and M. Nieto-Vesperinas, Opt. Lett.  27
(2002) 885.

\bibitem{Pendryprl} J.B. Pendry, Phys. Rev. Lett. 85 (2001)
3966.

\bibitem{Zhang} Z.M. Zhang, C.J. Fu, Appl. Phys. Lett. 80 (2002)
1097.

\bibitem{Jianqi} J.Q. Shen, Phys. Scr. 68 (2003) 87.

\bibitem{Shen}  J.Q. Shen, cond-mat/0308349 (2003).

\bibitem{Pendry2} J.B. Pendry, A.J. Holden, D.J. Robbins, and
W.J. Stewart, J. Phys. Condens. Matter  10 (1998) 4785.

\bibitem{Pendry1} J.B. Pendry, A.J. Holden, W.J. Stewart, and
I. Youngs, Phys. Rev. Lett.  76 (1996) 4773; J.B. Pendry, A.J.
Holden, D.J. Robbins, and W.J. Stewart, IEEE Trans. Microwave
Theory Tech.  47 (1999) 2075.


\bibitem{Maslovski} S.I. Maslovski, S.A. Tretyakov, and P.A.
Belov, Inc. Microwave Opt. Tech. Lett. 35 (2001) 47.

\bibitem{Smith2} D.R. Smith and N. Kroll, Phys. Rev. Lett. 85 (2000) 2933.

\end{references}
\end{document}